# An evaluation of the osmotic method of controlling suction

Pierre Delage, Yu Jun Cui

Ecole Nationale des Ponts et Chaussées, Paris, France

CERMES (department of Geotechnical engineering) – Institut Navier

*ENPC-CERMES, 6-8 Av. B. Pascal, F 77455 Marne la Vallée cedex 2 - France*
*Corresponding author: Pierre Delage*
*delage@cermes.enpc.fr*
*Tel. 33 1 64 15 35 42, fax 33 1 64 15 35 42*

**Abstract :** Experimental techniques of testing the mechanical properties of unsaturated soils are complex and difficult to conduct. As a consequence, complete sets of parameters that characterise the behaviour of unsaturated soils remain scarce and necessary. In this context, it has been found useful to gather the information obtained after some years of practice of the osmotic technique of controlling suction. As compared to the more documented axis-translation technique, the osmotic technique has its own advantages and drawbacks that are discussed in this paper, together with some potential future developments.

The osmotic method has been developed by soil scientists in the 1960s and adapted to geotechnical testing in the early 1970s. This paper presents the osmotic technique and comments on its advantages (including suction condition close to reality and higher suctions easily attained) and drawbacks (including some concern with the membrane resistance and some membrane effects in the suction/concentration calibration). Various applications to geotechnical testing are presented such as the determination of the water retention curve, oedometer and triaxial testing procedures and the determination of the permeability of unsaturated soils. Recent developments, that include the extension of the method up to high suctions (10 MPa) are also described, together with some recent and novel applications such as data from high controlled suction oedometer compression test and the determination of the oil/water retention properties of oil reservoir chalks.

1. Introduction

The behaviour of unsaturated soils is complex and still not fully understood in terms of mechanical response, fluid transfers and coupled hydro-mechanical behaviour. Proper understanding of the properties of unsaturated soils requires sophisticated experimental devices. As a consequence, complete sets of relevant experimental data remains scarce. However, significant progress has been made in constitutive and numerical modelling of problems related to unsaturated soils based on limited experimental data.

The method of controlling suction that is most often used when investigating the mechanical and hydraulic properties of unsaturated soils in a range of suctions generally comprised between 0 and 1500 kPa is the axis-translation technique (Fredlund and Rahardjo 1993). This technique consists in applying to the sample an air overpressure, whereas the water pressure is maintained constant (generally equal to atmospheric pressure). This technique (Hilf 1956) is operated by using a pressure plate apparatus (Richards 1941) where a ceramic porous stone of fine porosity enables the distinct control of air and water pressures. The first application of the axis translation method to geotechnical testing was the suction controlled triaxial apparatus developed by Bishop and Donald (1961) that is still used.

An alternative method of controlling suction called the osmotic technique was introduced later on in geotechnical engineering by Kassiff and Ben Shalom (1971), who proposed a novel suction-controlled oedometer (Figure 16). In this method, the soil sample was placed in contact with a semi-permeable membrane behind which an aqueous solution of large sized molecules of polyethylene glycol (PEG) was circulated. The circulation was made by using a burette connected to the base and the piston of the oedometer. This resulted in the application of a matrix[1] suction to the soil, that increased with the PEG concentration. The method was afterwards adapted to triaxial testing on hollow cylinder samples by Livneh et al. (1981) and on standard samples by Delage et al. (1987) and Cui and Delage (1996).

The method has been used and improved by various research groups over the world (Fleureau et al. 1993, Dineen and Burland 1995, Slatter et al. 2000a, b, 2005 and 2007, Tarantino and Mongiovi 2000, Cuisinier and Masrouri 2005, Monroy et al. 2007) in such a way that it has been considered useful to propose in this paper a synthetic description of the method, of the various developments carried out and to make some comments about its advantages and drawbacks as compared to other techniques of controlling suction.

2. Materials and methods

The semi-permeable membranes most often used in the osmotic technique are dialysis membranes used in medicinal field. This technique was used to control the osmotic pressure of culture solutions in biology by Lagerwerff et al. (1961) and it was applied later to the control of soil water matrix potential in soil science by Painter (1966), Zur (1966) and Waldron and Manbeian (1970). Peck and Rabbidge (1969) designed an osmotic tensiometer for measuring suction and the first application to geotechnical engineering was by Kassiff and Ben Shalom (1971).

The polyethyleneglycol (PEG) is a polymer made up of large sized molecules consisting in long chains with the following formula : HO-[$CH_2$-$CH_2$-O]$_n$-H. According to the value of $n$ that characterises the length of the chain, the molecular weight of the PEG changes.

---

[1] Note that the *osmotic* technique controls a *matrix* suction

Commercially available PEG have molecular weights included between 1000 and 50 000. PEG 6 000 and 20 000 are most commonly used in geotechnical testing.

The semi-permeable membranes are most commonly cellulotic membranes manufactured from natural cellulose reconstituted from cotton fibres. They are composed of fibres that form a grid that does not allow the passing through of large molecules. In the work carried out at ENPC-CERMES, Spectrapor[2] membranes were used. Membrane are provided in a dry state, impregnated by a preservative (glycerine) and they must be previously soaked in distilled water for at least 30 minutes before use. To fit with the various sizes of the PEG molecules, the semi-permeable membranes also have various degrees of fineness, defined by their molecular weight cut off (MWCO). Each PEG solution must be used with the corresponding MWCO, as shown in Table 1. Williams and Shaykewich (1969) reported that MWCO 3 500 semi-permeable membranes have a mean pore diameter of approximately 24 Å (2.4 nm) to be compared with the $\approx 1$ Å length of the OH link in the water molecule. An alternative to cellulotic membranes are polyether sulphonated synthetic membrane as initially proposed by Slatter et al. (2000a) and followed by Monroy et al. (2007).

The impedance of the membrane ($I_m = e_m / K_m$, where $e_m$ and $K_m$ are respectively the thickness membrane and permeability) is an important parameter that controls the time to reach suction equilibration. This aspect is poorly documented in the literature dealing with the osmotic technique. Impedance effects due to ceramic porous stones were commented in details in the description of a transient method of determining the permeability of unsaturated soils (Miller and Elrick 1958, Kunze and Kirkham 1962) and in the calculation of the relevant rate of testing of unsaturated soils in the suction controlled triaxial apparatus (Ho and Fredlund 1982). Semi-permeable membranes are much thinner than ceramic porous stones that they sometimes replace in the axis-translation method. Painter (1966) gives a thickness value $e_m = 90$ µm for a MWCO 4 500 Visking dialysis membrane and Suraj De Silva (1987) determined a thickness of 47.5 µm for a MWCO 12 000-14 000 Spectrapor 2 wet membrane.

Various variable head permeability tests carried out by Suraj De Silva (1987) and Lepilleur and Schmitt (1995) on semi-permeable membranes placed in a oedometer showed that the coefficient of permeability $k_m$ of the Spectrapor 2 membrane (12 000-14 000 MWCO) was equal to $10^{-12}$ m/s, as compared to $1.3 \times 10^{-13}$ m/s for Spectrapor 1 (3500 MWCO, same thickness). This is compatible with the results obtained by Slatter et al. (2000a) on both a cellulotic (MWCO 14 000, $1 \times 10^{-13} < k_m < 3 \times 10^{-13}$ m/s) and a polyether sulphonated membrane ($k_m = 1.5 \times 10^{-13}$ m/s). Dineen (1997) in Monroy et al. (2007) reports values of 1.5 to $2.9 \times 10^{-13}$ m/s for a 12 000 – 14 000 MWCO cellulose membrane whereas Monroy et al. (2007) give a value of $8 \times 10^{-11}$ m/s for a polyether sulphonated membrane, significantly smaller than the value of $1.5 \times 10^{-13}$ m/s given by Slatter et al. (2000). These values are in the same order of magnitude as the coefficient of permeability of dense plastic clays. Obviously, higher MWCO membranes have larger pores and a larger coefficient of permeability, that makes preferable their use together with PEG 20 000 to reduce testing times. Semi-permeable membranes are thinner and are about one order of magnitude less permeable than ceramic porous stones ($k_p = 1.2 \times 10^{-9}$ and $8 \times 10^{-11}$ m/s for 5 and 15 bars air entry values (AEV) porous stones, according to Ho and Fredlund 1982). Due to the combined effect of the differences in coefficient of permeability and thickness, the impedance of a 12 000-14 000 MWCO membrane ($5 \times 10^7$ s) is close to that of a 15 bars 6 mm thick AEV porous stone ($7.5 \times 10^7$ s).

---

[2] Manufactured by Spectrum Co. Other brands are Visking, Millipore, etc...

An important problem related to the use of cellulotic semi-permeable membranes is their fragility. Cellulotic semi-permeable membranes are sensitive to mechanical shearing and bacteria attacks and their degradation may allow for PEG crossing and penetrating the sample. The sensitivity to bacteria attack can be corrected with the addition of a few drops of penicillin in the solution (Kassiff and Ben Shalom 1971). Experience showed that in this case, a cellulotic semi-permeable membrane could last up to 10 days (Suraj de Silva 1987), allowing for the completion of a suction controlled triaxial test in most cases. When possible, it seems however preferable to change a membrane after 6 days. In this regard, recent results by Monroy et al. (2007) showed that synthetic polyether sulphonated membranes have a better resistance, allowing to run tests as long as 146 days with no evidence of leaks. The mechanical resistance to shear of membranes will be commented later in a different section.

A convenient way of determining the concentration of a PEG solution consists in measuring the refraction degree of the solution, using a hand refractometer[3] (Lagerwerff et al. 1961, Suraj De Silva 1987). This measurement gives the Brix value of the solution (a 0% Brix value corresponds to pure water whereas a 100% Brix value is the refractive index of anhydrous saccharose). Figure 17 shows the calibration carried out by Delage et al. (1998) on four different PEGs of high concentrations, giving the concentration as a function of the refractive index of the solution.

3. Calibration of the method

Initially, the calibration curves giving the total suction as a function of the solution concentration of various PEGs were investigated by measuring the relative humidity above solutions of PEG by using psychrometers (Lagerwerff et al. 1961, Zur 1966).

Figure 18 presents the calibration results independently obtained on PEGs 6 000 and 20 000 by various authors and gathered by Williams and Shaykewich (1969). The data show no difference between points from PEG 6 000 and from PEG 20 000, the calibration curve being independent on the molecular mass of the PEG used. Based on the results of Figure 19 that presents the retention curves of two soils determined by using both the osmotic technique (using Williams and Shaykewich's calibration) and the pressure membrane technique, Zur (1966) considered that a satisfactory agreement was obtained between the two techniques. The figure shows a good agreement with the Natania sandy loam (good agreement was also observed by the author on the Columbia fine sandy loam) and with some minor differences on the Aylon clay, with smaller water contents obtained at higher suction (s > 400 kPa) with the osmotic technique. The difference observed in the figure between the data obtained with the two techniques could be related to a too short period of equilibration time when using the osmotic technique (72 hours). A better correspondence might be expected with longer equilibration period (see the data of Figure 24 and the related comments further on). Waldron and Manbeian (1970) drew similar conclusions on 6 California soils. The calibration curve of Figure 18 has been used by various researchers including Kassiff and Ben Shalom (1971), Delage et al. (1987, 1992), Fleureau et al. (1993) and Cui and Delage (1996). As discussed earlier, a calibration based on psychrometer measurements gives, through the equilibrium in the vapour phase, a measurement of the total suction, whereas the osmotic technique using a membrane controls, through the liquid phase, the matrix suction of the soil sample, the difference between the two being related to the osmotic component of the suction.

Dineen and Burland (1995) used a newly developed tensiometer (Ridley and Burland 1993) able to measure suctions up to 1 500 kPa to further investigate the calibration curves of PEG. They made direct suction measurements on a sample kept under a suction controlled by the osmotic technique in a oedometer, allowing the direct calibration of the matrix suction of the sample placed in contact with the semi-permeable membrane. Dineen and Burland (1995) also developed a special device to measure directly the suction by placing the probe in contact, through a kaolinite thin layer, with the semi-permeable membrane behind which the solution was circulated. As seen in Figure 20, Dineen and Burland evidenced that the membrane had an effect on the value of the suction imposed to the sample by the osmotic technique, resulting in a smaller value than that obtained from psychrometer measurements. Note also the good agreement between the direct probe measurements made in contact with the semi-permeable membrane and the measurements obtained on kaolin samples in the osmotically suction controlled oedometer. The figure shows that the membrane effect begins to be noticeable at suction higher than 200 kPa. As compared to the psychrometer calibration, the membrane effect reduces the suction of less than 100 kPa in the zone of suctions of 500 kPa. Around 1 000 kPa, the reduction is approximately 200 kPa. Actually, a similar membrane effect had also been observed from the data of Waldron and Manbeian (1970) who developed a null type osmometer in which the osmotic pressure was compensated by an air pressure applied to the solution for suctions included between 16 and 2480 kPa. The corresponding points are also reported in Figure 20, they are located on Dineen and Burland's curve (except the point at 1500 kPa), providing an extension of the calibration up to higher suctions (2480 kPa).

---

[3] An ATAGO refractometer was used in this study

Dineen and Burland also referred to the data obtained by Peck and Rabbidge (1969) also represented in the figure. Peck and Rabbidge's used an osmotic tensiometer and their data give even a smaller suction for the same concentration. However, some specific problems related to the use and accuracy of the osmotic tensiometer commented by the authors themselves and also by Fredlund and Rahardjo (1993) have to be recalled when considering these data. Slatter et al. (2000) also investigated some possible membranes effect in PEG calibration. They used an osmotic oedometer and developed an osmotic pressure cell. They also observed smaller suctions at the same PEG concentrations when using a membrane. The calibration points obtained with cellulose acetate membranes are shown in Figure 20. These results are closer to those of Peck and Rabbidge (1963) for suctions higher than 0.3 MPa. This is not surprising since the principle of the osmotic pressure cell and that of the osmotic tensiometer are close. To account for this effect, they proposed a thermodynamic analysis where the difference in suction due to the use of a membrane is related to an increase of entropy related to a possible organisation of PEG molecules due to surface interactions at the contact with the membrane.

Tarantino and Mongiovi (2000) and, more recently, Monroy et al. (2007), performed tests similar to that of Dineen and Burland (1995). They evidenced that calibration was dependent of the nature of the membrane and PEG used, as seen in Figure 21 in which the following systems were used: i) MWCO 14 000 cellulotic membrane (brand not given) and PEG 20 000 (Dineen and Burland 1995); ii) MWCO 14 000 cellulotic Spectrum membrane and PEG 20 000 (Tarantino and Mongiovi 2000), iii) MWCO 14 000 cellulotic Viskase membrane and PEG 20 000 (Tarantino and Mongiovi 2000) and iv) MWCO 15 000 polyether sulfone membrane and PEG 35 000 (Monroy et al. 2007). In accordance with Slatter et al. (2000), Monroy et al. (2007) observed that, for a given concentration, the highest suctions were obtained by using the polyether sulfone membrane with PEG 35 000 with suction values close to that of Williams and Shaykewich (1969). Monroy et al. (2007) also observed some difference when comparing calibration points along a wetting path (suction decrease) compared to that along a drying path (suction increase) with smaller suction obtained during the subsequent drying path.

## 4. Extension to higher suctions

Soil scientists and geotechnical engineers generally use the osmotic technique up to a 1 500 kPa suction, except Waldron and Manbeian (1970) who worked up to 2 480 kPa with PEG 6 000. The necessity to deal with higher suctions when investigating the behaviour of engineered barriers made up of compacted swelling soils for the isolation of radioactive waste disposal led to the extension of the method. It was found that higher suctions could be reached with smaller PEG molecules by increasing the PEG concentration closer to the saturated concentration of the solution (Delage et al. 1998). The calibration curve at higher suction was determined by placing containers of PEG solutions at higher concentrations in desiccators containing saturated salt solutions at a controlled temperature (20°C). Equilibration was checked by weighing regularly the PEG containers until weight stabilisation that was obtained after three weeks.

Table 2 presents the salt used for this calibration, the corresponding relative humidity (between 91.3 and 97%) and the suctions obtained together with the corresponding PEG concentration at equilibrium for PEG 20 000, 6 000, 4 000 and 1 500 (see Delage et al. 1998 for more details). The occurrence of precipitation is also mentioned. The table confirms that precipitation occurs at lower PEG concentration with larger molecular mass and shows that maximum suctions of 9 and 12.6 MPa can be respectively attained with PEG 4 000 and 1 500.

It was found relevant to plot the data in a $\sqrt{suction}\,/\,concentration$ diagram. Figure 22 shows the data plotted together with the points gathered by Williams and Shaykewich (1969). A good agreement is observed between all data, with no dependence on the molecular weight, showing that good confidence can be given to the calibration. This presentation also shows that the following empirical relation is valid at suctions smaller than 4 MPa :

$$s = 11\,c^2 \tag{1}$$

where *s* is expressed in MPa and *c* in g PEG / g water.

The calibration curve of Figure 22 does not account for the membrane effect previously described and a further calibration including the membrane effect should be carried out. Since the method of Dineen and Burland could not be used at suctions higher than 1 500 kPa, it seems that an extension to high pressures of the null type osmometer used by Waldron and Manbeian (1971) could provide an extension of the calibration curve accounting for membrane effects. If the membrane effect appeared not to be suction dependent and to correspond to a 200 kPa decrease in suction as evidenced in Figure 20, the corresponding relative error would obviously decreases at higher suction. This has to be confirmed and further experimental investigation is obviously needed in this regards.

The extension of the osmotic technique up to 10 MPa is believed to be of interest as compared to the extension of the axis translation method up to a similar value (Escario and Juca 1989), that requires heavy equipment able to support high air pressure in safe conditions. In this range of suctions, the osmotic technique is considerably simpler, safer and inexpensive and it would allow the continuity of the techniques of controlling suction between lower suctions (< 1 500 kPa) and higher suctions (various tens of MPa controlled with vapour equilibrium).

Up to 10 MPa, the extension of the osmotic technique could also provide an alternative method to the vapour equilibrium technique, that has some drawbacks at lower suctions, i.e. a lower accuracy and significantly longer equilibration times. Since water exchanges occur in the liquid phase in the osmotic technique, equilibration times are much shorter. The fact that the osmotic technique controls the matrix suction whereas the vapour equilibrium technique controls the total suction may help in further investigating the osmotic component of the suction in this range.

## 5. Applications in geotechnical engineering

As compared to the axis translation method, the main advantage of the osmotic technique is that no artificial air pressure ($u_a > 0$) is used to apply a suction ($s = u_a - u_w > 0$) to the soil, like in reality (Kassiff and Ben Shalom 1971). The suction is indeed applied in a zero air pressure condition ($u_a = 0$) with a "negative" water pressure condition ($u_w < 0$) exerted by the osmosis phenomenon through the membrane, up to high values if necessary. This difference is likely to be of interest when investigating higher degrees of saturation near full saturation, when the air phase becomes discontinuous and where some limitations of the air overpressure method have been described (Bocking and Fredlund 1980). Obviously, a comparative investigation of the water retention and permeability properties of various typical soils carried out by using both the osmotic and axis-translation methods in the range of high degrees of saturation should provide novel and useful results.

As commented in Zur (1966) and Waldron and Manbeian (1970) (see also Cui and Delage 1996 and Delage et al. 2001), the use of the osmotic technique for the determination of the water retention curve is simple and inexpensive, since no air pressure device is necessary. As shown in Figure 23 for a triaxial sample, the sample is introduced in a semi-permeable membrane tubing that is plunged in a container full of the PEG solution and placed on a magnetic stirrer. A good contact must be ensured between the sample and the membrane by using an adapted semi-permeable tubing and by placing O-rings around the sample at various levels. To avoid water evaporation, it is recommended to use a plastic film to isolate the solution from ambient relative humidity. The period of time necessary to reach equilibrium depends on the size of the sample and of the permeability of the membrane. For 90 mm thick samples, Zur (1966) observed that a 48 hours period of time was too short. Figure 24 (Suraj de Silva 1987) shows the mass changes with time of 3 triaxial samples of compacted Jossigny silt (76 x 38 mm) submitted to a suction of 800 kPa using PEG 20 000. Samples were periodically withdrawn and the mass determined. The figure shows that a period of time of 20 days is necessary to reach equilibrium. During this period, the concentration of the surrounding solution decreases and the final value of the concentration (and of the corresponding suction) was determined using the refractometer. Some additional quantities of PEG may be necessary to achieve the desired concentration.

The system of Kassiff and Ben Shalom presented in Figure 16 was further improved by Delage et al. (1992), who replaced the burette by a closed circuit in which the solution was circulated by a peristaltic pump. As seen in Figure 25, the closed circuit also comprises a 1 litre bottle placed in a temperature controlled bath. The large volume of the bottle allows for a relatively constant value of the concentration in spite of the exchanges occurring through the membrane between the solution and the soil. A capillary tube allows for the monitoring of water exchanges during the tests and another one is used for controlling evaporation. The circulating system with the peristaltic pump of Delage et al. (1992) has been adopted by Dineen and Burland (1995) who proposed to place the bottle on an electronic balance in order to automatically record the changes in weight of the bottle, with no influence of temperature on the measurement system. Tarantino and Mongiovi (2000) and Monroy et al. (2007) also used this device.

The first application to triaxial testing was by Livneh et al. (1981) who developed a hollow cylinder triaxial sample (38 mm diameter and 76 mm high) and placed the membrane in the inner cylinder. The PEG solution was circulated inside the inner cylinder, leading to two problems: i) the solution had to be pressurised at the cell pressure and ii) the semi-permeable membrane, being permeable to water, could not properly apply the cell pressure. Delage et al. (1987) proposed a new triaxial for unsaturated soils by applying the same system as Kassiff and Ben Shalom (1971), i.e. by circulating the solution behind semi-permeable membranes placed on the piston and the base of the triaxial sample (see also Cui and Delage 1996).

As compared to the axis-translation method, the osmotic system is relatively simple and easier to adapt to the oedometer. No air tightness device is to ensure, leading to systems with less friction effects. Also, the membrane is only submitted to a one dimensional compression effort and experiments showed that it behaved correctly up to 1 600 kPa (Delage et al. 1992). As seen previously, high suction can be reached much more easily in the osmotic oedometer than by applying an air overpressure.

The use of the technique in triaxial testing is not as easy as the axis-translation method, where it is sufficient to replace the lower porous stone by a ceramic porous stone (Bishop and Donald 1962). To adapt the osmotic technique to the triaxial apparatus, Delage et al. (1987) used to glue the membrane on the triaxial base and piston using O-rings and an epoxy resin, which appeared to be a somewhat delicate operation. Actually, clamping systems on the base and the piston seem preferable. Another difficulty was encountered to ensure the $u_a = 0$ condition through an air vent that was machined in the centre of the lower base of the cell that required a delicate gluing of the membrane around. Actually, an alternative would consist in connecting an air pressure flexible duct through the membrane at the middle height of the sample. This was suggested by Maâtouk et al. (1995) to reduce by a factor of two the length of drainage when using the axis-translation technique. This option imposes the use of local strain measurements to monitor the volumetric strain.

The mechanical conditions applied to the semi-permeable membrane in the triaxial apparatus are more critical than in the oedometer, since the membrane is submitted to an extension stress. However, a Spectrapor 2 12-14 000 membrane exhibited a satisfactory resistance in a test carried out on a compacted Jossigny silt at a cell pressure of 400 kPa and a maximum deviatoric stress of 1300 kPa under a 1500 kPa controlled suction (Cui and Delage 1996).

As compared to these drawbacks, an advantage of the technique is that higher suction can be reached more easily, since there is no need to simultaneously increase both the cell pressure and the air pressure to keep a constant normal net stress ($\sigma$ - $u_a$) while increasing the suction ($u_a - u_w$) by increasing the air pressure $u_a$. In this regard, the maximum suction ever applied to suction controlled triaxial tests (1500 kPa) was applied using the osmotic technique (Cui and Delage 1996). Also, the presence of a membrane on top and bottom of the sample reduces the length of drainage by two, as compared to most common air overpressure triaxial devices.

As described in Delage et al. (1992), the osmotic oedometer can also be used for determining the coefficient of permeability of an unsaturated soil by using Gardner's method (Gardner 1956). In this method, an instantaneous increment of suction is applied to the sample at $t = 0$ and the quantity of fluid $Q(t)$ expelled so as to achieve suction equilibration is carefully monitored as a function of time. Based on a simplified resolution of Richards's equation, Gardner demonstrated that the logarithm of the water outflow was linear as a function of time according to the following relation:

$$Log[Q_0 - Q(t)] = Log \frac{8Q_0}{\pi^2} - \frac{\pi^2}{4L^2} Dt \qquad (2)$$

where $Q_o$ is the total outflow (in terms of volumetric water content), $D$ the water diffusivity (m$^2$/s) and $L$ the length of the sample.

In the osmotic technique, the suction is increased by changing the concentration in the bottle and monitoring either the change of solution level in the capillary tube or the change in weight of the PEG bottle. Figure 26 (Vicol 1990, Delage et al. 1992) presents the data obtained on a slurry made up of Jossigny silt under a 50 kPa load, when suction was increased from 100 to 200 kPa. As seen in the figure, the water outflow plotted in a Log Q vs time diagram is indeed linear, confirming Gardner's approach. The coefficient of permeability is calculated from the value of the diffusivity $D$ obtained from the slope of the curve.

## 6. Comparison with other techniques

A comparison between the osmotic technique and other methods of controlling suction (tensiometric plate, air overpressure, vapour equilibrium and thermocouple measurements) was carried out by Fleureau et al. (1993) on a kaolinite slurry on a wide range of suctions (0.4 up to 180 MPa. The calibration curve used was that of Williams and Shaykewich (1969) with no consideration of any membrane effect. Figure 27 presents the results in terms of void ratio vs suction. A good overall agreement between all techniques is observed in the drying stage that occurs in a saturated state until the air entry value (close to 2 MPa). Between 100 and 1500 kPa, excellent agreement is observed between air-overpressure, osmotic and thermocouple psychrometer data, showing a negligible effect of the osmotic component of the suction in this soil. In the light of Dineen and Burland (1995)'s calibration, the point at $s = 500$ kPa should be reduced of about 100 kPa and that at 1 000 kPa of 200 kPa. As seen in the figure, this correction would improve the correspondence between the two techniques. Note that the semi-log plot is somewhat hiding the membrane effect. Some problems between the axis translation method and the osmotic method appear in the re-wetting path between 1500 and 100 kPa, possibly close to the zone where air bubbles appear. This could be related to the problems of the axis translation method in the zone of occluded air (Bocking and Fredlund 1980, Fredlund and Rahardjo 1993). Further investigation on this point is obviously needed.

Figure 28 shows the water content vs suction relationship obtained on samples of an interstratified illite-smectite swelling clay considered as a possible engineered barrier for nuclear waste disposal at great depth called Fourges clay ($w_P = 50$, $w_L = 112$, Delage et al. 1998, Tessier et al. 1998). Data obtained on a powder and on a compacted sample using both the vapour equilibrium technique and the extended osmotic technique (psychrometer calibration) are compared. Here also, a reasonably good agreement is observed between the two techniques between 2 and 6 MPa suction, that should be confirmed by a calibration of the membrane effects in this suction range (although some significant dispersion is observed at 6 MPa on the powder samples with the vapour equilibrium technique). A singular property of these very active clays is that there is no significant difference between the powder and compacted samples. Also, the responses to suction cycles appeared to be reversible (Delage et al. 1998). These two aspects show that the physico-chemical clay-water interactions, that are known to act reversibly (Gens and Alonso 1992, Delage et al. 1998) play a dominant role in the retention properties of the soil, as compared to the standard hysteretic capillary effects that govern water retention in inactive porous media.

Figure 29 (Cuisinier and Masrouri 2001) shows the results of two constant suction compression oedometer tests carried out on a mix of silt (40%) and commercially available bentonite (60%) at a suction of 8.5 MPa using both the vapour equilibrium technique and the extended osmotic technique (psychrometer calibration). The two curves are comparable with similar slopes in the plastic zone and a smaller yield stress in the test with vapour control, showing the potentialities of both methods for testing under high controlled suction.

Figure 30 presents a novel application of the osmotic technique in oil engineering, i.e. the oil retention curve of a reservoir chalk (Priol et al. 2004). The figure presents the wetting path (osmotic technique) and the drying path (axis-translation method) of the curve that are also compared to the drainage curve that can be derived from mercury intrusion porosimetry. Samples of chalk (Lixhe chalk from Belgium, porosity $n = 45\text{-}50\%$) full of hydrocarbon (Soltrol 170) were placed in semi-permeable tubing immersed in the PEG solutions at various concentration (see Figure 23). Progressively, water infiltrated at a controlled suction within the sample and expelled oil. The oil/water suction $u_o$ - $u_w$ is controlled through the changes of $u_w$, $u_o$ being equal to atmospheric pressure. During the experiment, the membrane was not affected by the contact with oil. The shape of the water wetting curve confirms the unimodal pore size distribution defined by the $\approx 1\mu m$ diameter coccoliths, and most water expels the oil below a 500 kPa oil-water suction, giving a residual degree of saturation in water of 70%. The ood correspondence between the drainage paths from the axis-translation technique and from mercury intrusion porosimetry show the predominant effect of capillarity in the retention phenomena. Water retention properties of unsaturated chalk samples from a chalk abandoned quarry have obtained in a similar manner been presented in De Gennaro et al. (2006).

## 7. Conclusions

A description of the osmotic method of controlling matrix suction was proposed as a possible alternative to the standard axis-translation method and the advantages and drawbacks of the method were discussed.

The concentration versus suction calibration curves obtained in the literature using psychrometer measurements showed that compatible results were obtained by various authors, independently of the molecular mass of the PEG. By performing direct matrix suction measurement on samples that were suction controlled using the osmotic technique, Dineen and Burland (1995) obtained a different calibration curve, evidencing an effect due to the membrane that results in smaller suctions. There results are in agreement with results obtained by Waldron and Manbeian (1970) using a null type osmometer. Membrane effects were also further confirmed by Tarantino and Mongiovi (2000), Slatter et al. (2000b) and Monroy et al. (2007) who showed that the calibration was also dependent of the type of membrane used.

Obviously, the condition of suction applied by the osmotic technique is closer to reality than in the axis-translation method since no artificial pressure is applied to air. Also, the recent extension of the osmotic technique up to suctions as high as 10 MPa seems of a potential significant interest in the experimental investigation of unsaturated soils. Since this extension was made by controlling the suction through the control of the relative humidity, membrane effects should be investigated at higher suction. An adaptation to high pressures of the null type osmometer of Waldron and Manbeian (1971) seems a possible way to account for membrane effects at higher suctions.

A well known draw-back of the method is the fragility of cellulotic membranes to bacteria attack and stresses. The use of penicillin in the solution appeared to provide satisfactory membrane resistance up to 2 weeks. Another interesting option is the use of synthetic polyether sulfonated membrane as suggested by Slatter (2000a). Polyether sulfonated membranes allow to run tests along a duration period of times up to several months (Monroy et al. 2007). In the oedometer where no extension stress is applied to the membrane due to the lateral confinement of strain, satisfactory behaviour was observed in the stress range of standard geotechnical engineering. The situation is more critical in the triaxial apparatus where some extension effort may induce the shearing of the membrane. This problem was solved by using a membrane described as more resistant by the manufacturer, that appeared to behave satisfactorily under a deviator stress of 1 300kPa.

In the oedometer, the adaptation of the osmotic technique is simple and allows for the easy application of high suction without air-tightness and friction problems. In the triaxial, the use of clamping systems on base and piston with the connection of the air vent through the membrane at mid-height is probably the easiest way to reach significantly high suction with reasonable simplicity. The osmotic technique also allows simple determination of the coefficient of permeability of unsaturated soils and is probably particularly interesting for low permeability values (in the same range as the membrane permeability i.e. $10^{-12}$ m/s) in dense clay soils, since high suction gradient may be easily applied. Recent results also showed that the osmotic technique could be used in petroleum engineering to investigate the oil-water system in reservoir porous rocks.

## 9. Acknowledgements

The contribution of the various PhD students who worked at ENPC using the techniques described in this paper is gratefully acknowledged: Drs G.P.R. Suraj de Silva (1987), T. Vicol (1990), M. Yahia-Aissa (1999), Marcial (2003) and Priol (2005). The collaboration of M. Howat[4], E. De Laure and V. De Gennaro was also greatly appreciated.

---

[4] Deceased

10. Tables

Table 1 : PEG solutions and corresponding semi-permeable membranes (defined by their MWCO)

| PEG Solution | Molecular weight cut off (MWCO) |
|---|---|
| 20 000 | 12 000 – 14 000 |
| 6 000 | 3 500 |
| 4 000 | 2 000 |
| 1 500 | 1 000 |

Table 2 : PEG Calibration at high concentration (Delage et al. 1998)

| Salt | Hr (%) | Suction (MPa) | Concentration PEG 20 000 (g PEG / g water) | Concentration PEG 6 000 (g PEG / g water) | Concentration PEG 4 000 (g PEG / g water) | Concentration PEG 1 500 (g PEG / g water) |
|---|---|---|---|---|---|---|
| $K_2SO_4$ | 97 | 4.2 | | | | 0.599 |
| $CuSO_4$ | 95.7 | 6.1 | | | 0.714 | |
| $KH_2PO_4$ | 95.5 | 6.3 | | 0.758 | | |
| $K_2NO_3$ | 93.7 | 9 | Precipitation | 0.990 | | 0.963 |
| $Na_2HPO_4$ | 93.7 | 9 | | | 0.952 | |
| $ZnSO_4$ | 91.3 | 12.6 | Precipitation | Precipitation | | 1.350 |

11. Figures

Captions to Figures

Figure 1 : The osmotic oedometer of Kassiff and Ben Shalom (1971)

Figure 2 : Calibration of the refractive indexes of various PEGs (1 500, 4 000, 6 000 20 000) at high concentrations (modified after Delage et al. 1998)

Figure 3 : Calibration curves of PEGs 6 000 and 20 000 obtained with psychrometer measurements of the relative humidity of the solutions (modified after Williams and Shaykewich 1969).

Figure 4 : Equilibrium moisture contents at various matrix suctions for osmotic system and pressure membrane (modified after Zur 1966)

Figure 5 : Suction/concentration calibration curve obtained by Dineen and Burland (1995), completed with data from Waldron and Manbeian (1970) and Slatter et al. (2000b)

Figure 6: Dependency of the calibration curve with respect to the membrane used

Figure 7 : Calibration curve at higher suction (modified after Delage et al. 1998)

Figure 8 : Osmotic device for the determination of the water retention curve (Cui and Delage 1996)

Figure 9 : Time to suction equilibration in triaxial samples submitted to a 800 kPa suction (Suraj De Silva 1987)

Figure 10 : Adaptation of a closed circuit to the osmotic oedometer (modified after Delage et al. 1992)

Figure 11: Adaptation of Gardner's method of measuring the permeability of unsaturated soils (Vicol 1990, Delage et al. 1992).

Figure 12 : Comparison between various suction controlled techniques (Fleureau et al. 1993)

Figure 13 : Water retention properties of a engineered barrier FoCa7 clay (Yahia-Aissa et al. 2000)

Figure 14 : High suction (8.5 MPa) oedometer compression tests (Cuisinier and Masrouri 2005)

Figure 15 : Oil retention properties of a chalk (Priol et al. 2004)

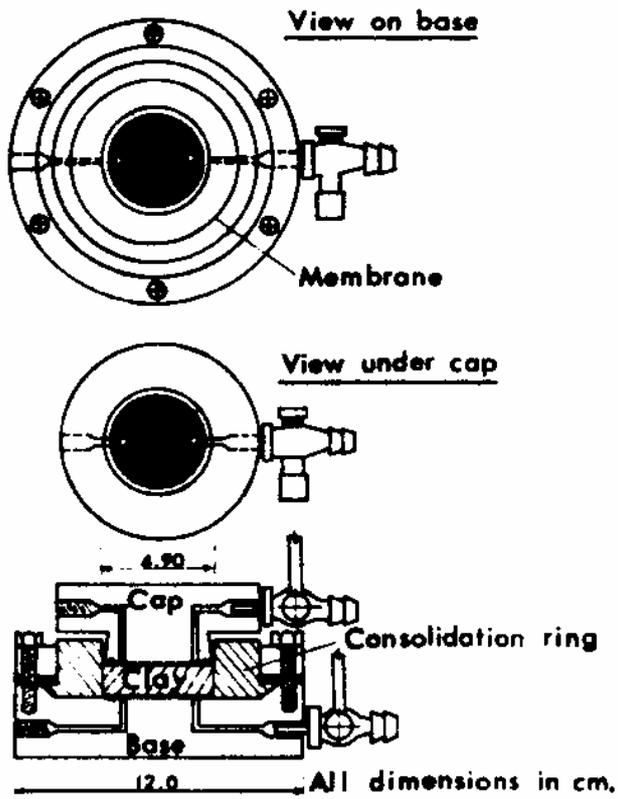 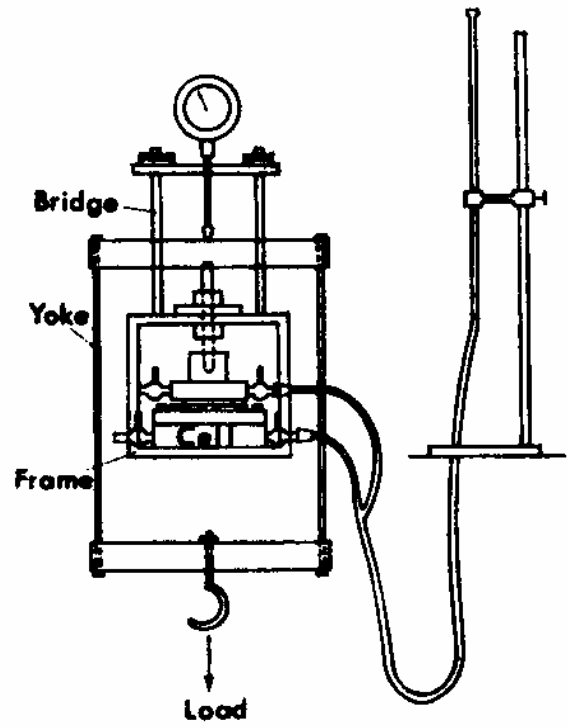

Figure 16 : The osmotic oedometer of Kassiff and Ben Shalom (1971)

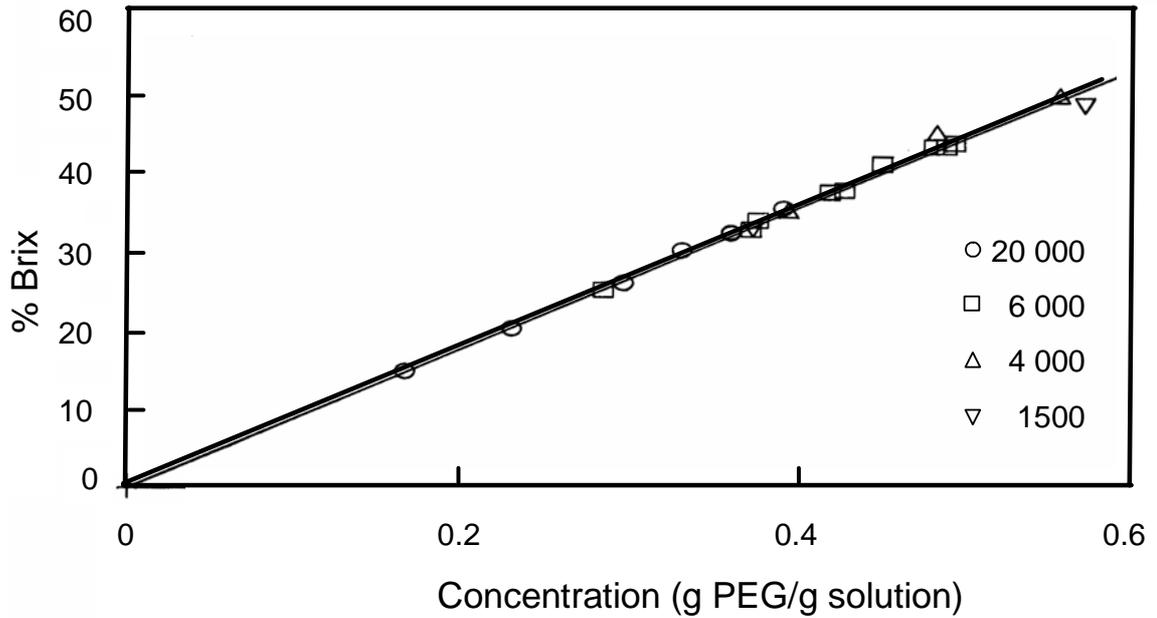

Figure 17 : Calibration of the refractive indexes of various PEGs (1 500, 4 000, 6 000 and 20 000) at high concentrations (modified after Delage et al. 1998)

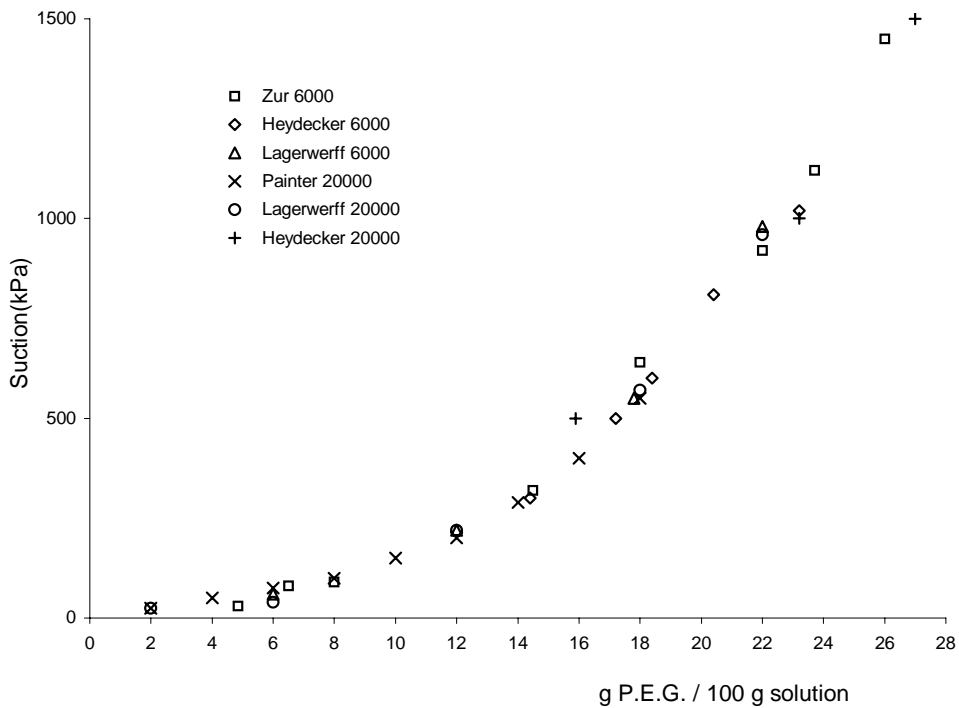

Figure 18 : Calibration curves of PEGs 6 000 and 20 000 obtained with psychrometer measurements of the relative humidity of the solutions (modified after Williams and Shaykewich 1969).

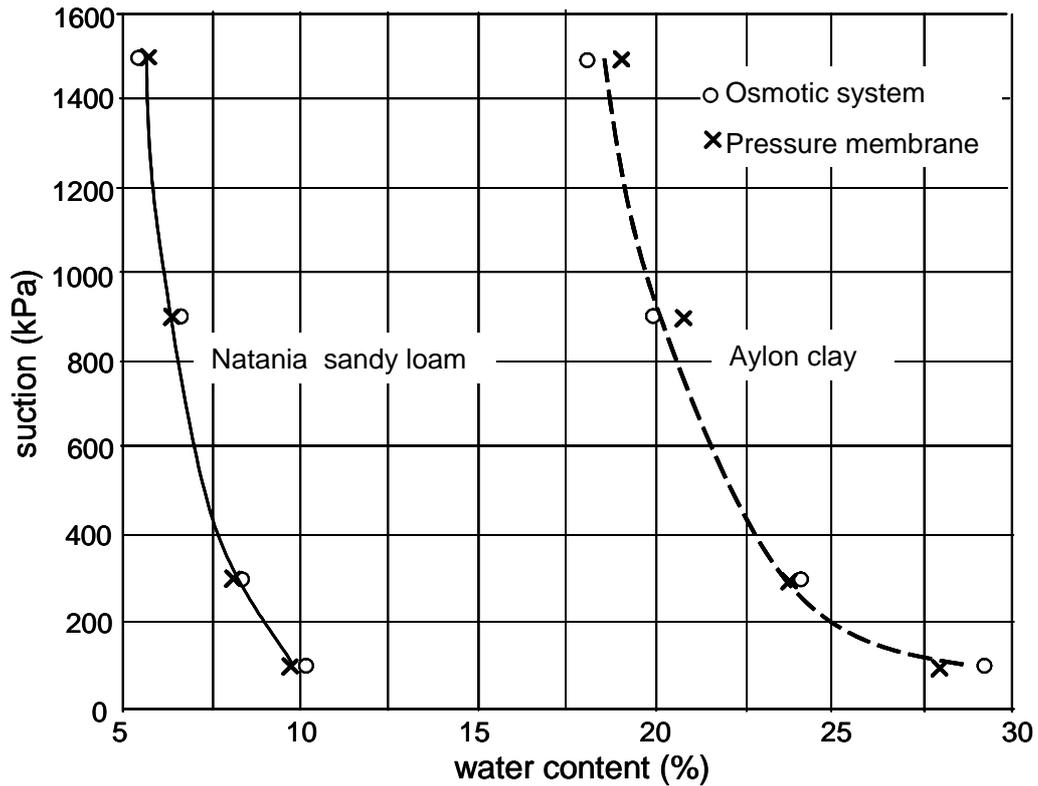

Figure 19 : Equilibrium moisture contents at various matrix suctions for osmotic system and pressure membrane (modified after Zur 1966)

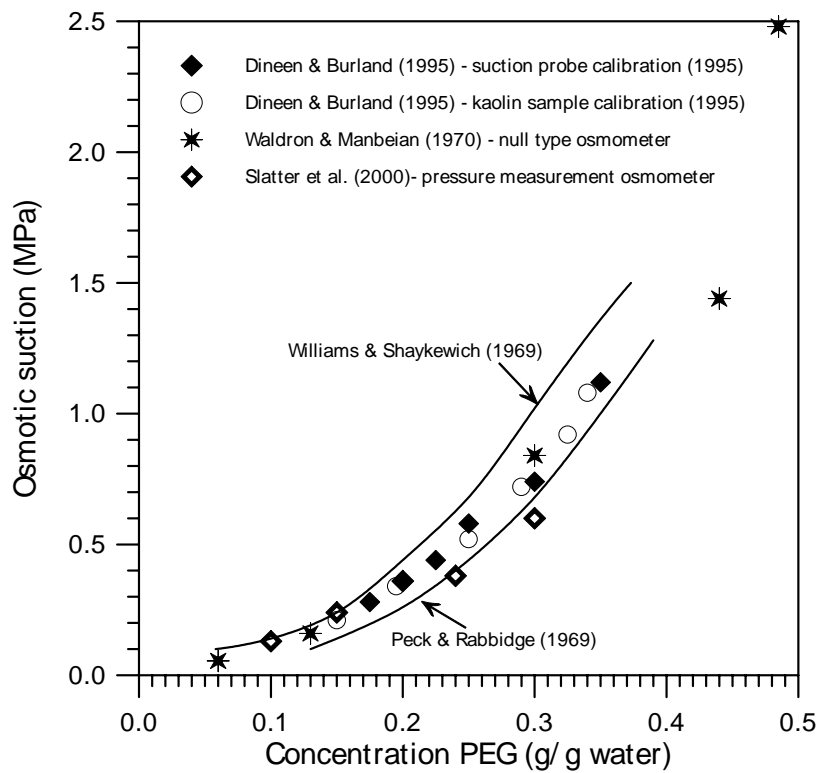

Figure 20 : Suction/concentration calibration curve obtained by Dineen and Burland (1995), completed with data from Waldron and Manbeian (1970) and Slatter et al. (2000b)

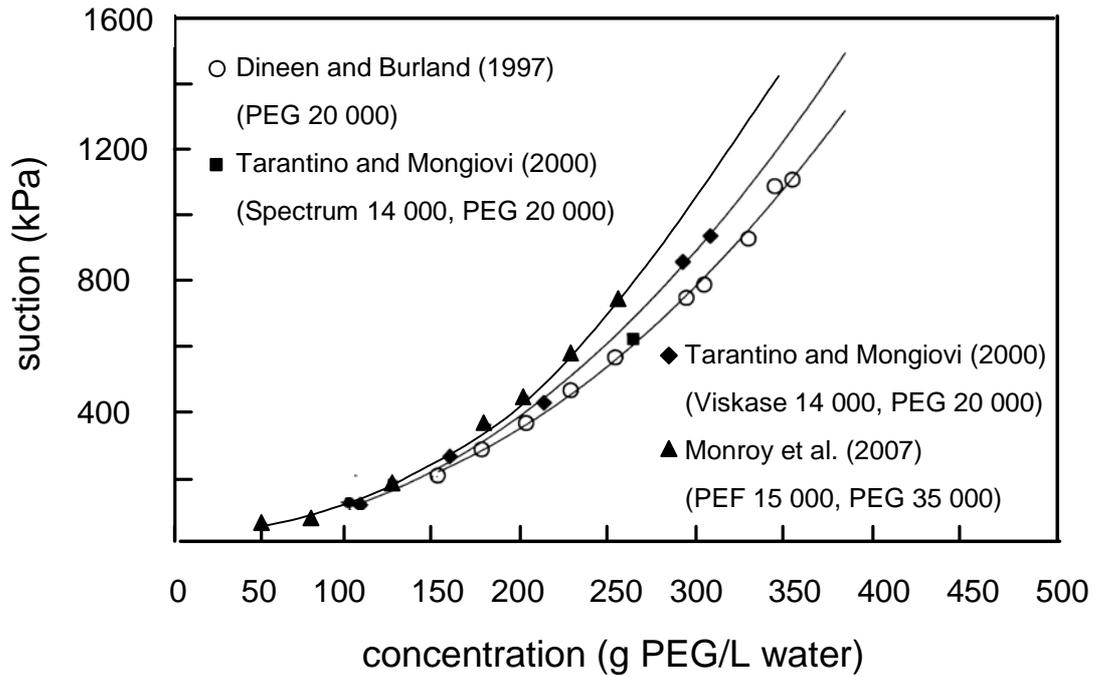

Figure 21: Dependency of the calibration curve with respect to the membrane used

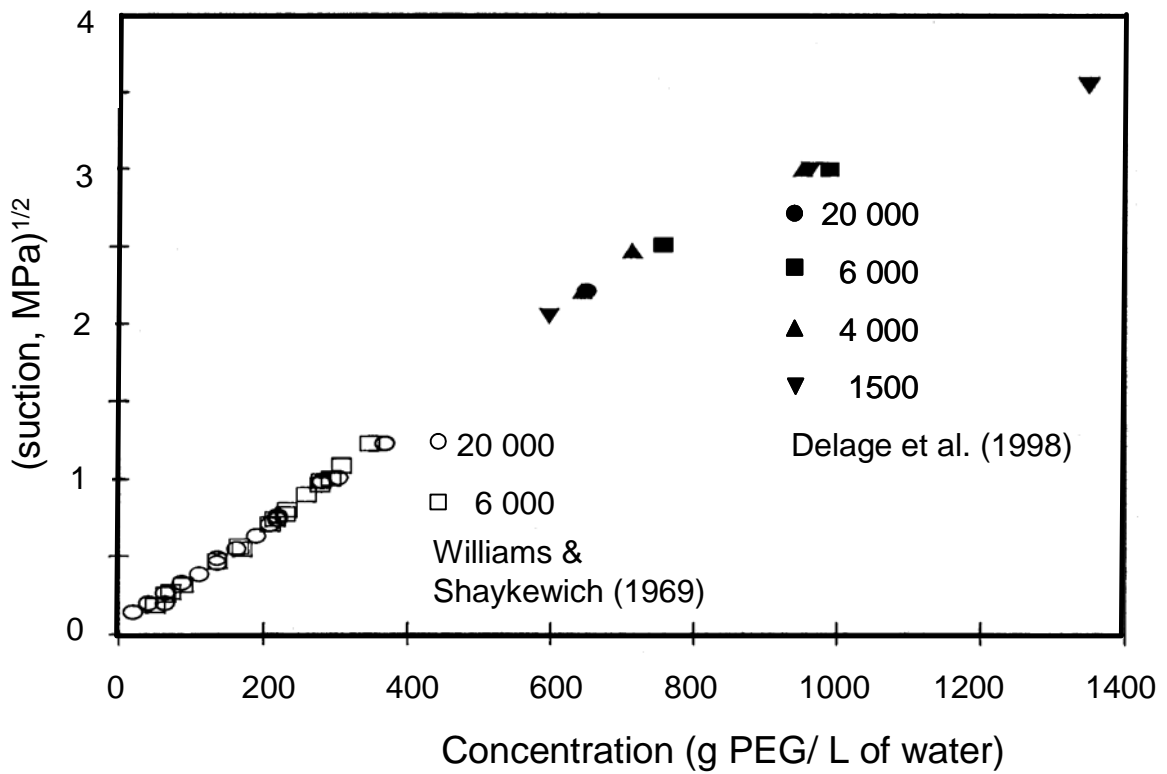

Figure 22 : Calibration curve at higher suction (modified after Delage et al. 1998)

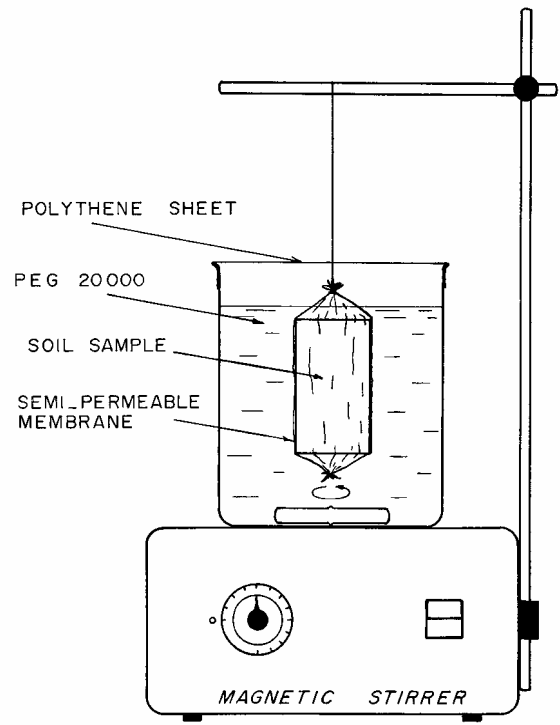

Figure 23 : Osmotic device for the determination of the water retention curve (Cui and Delage 1996)

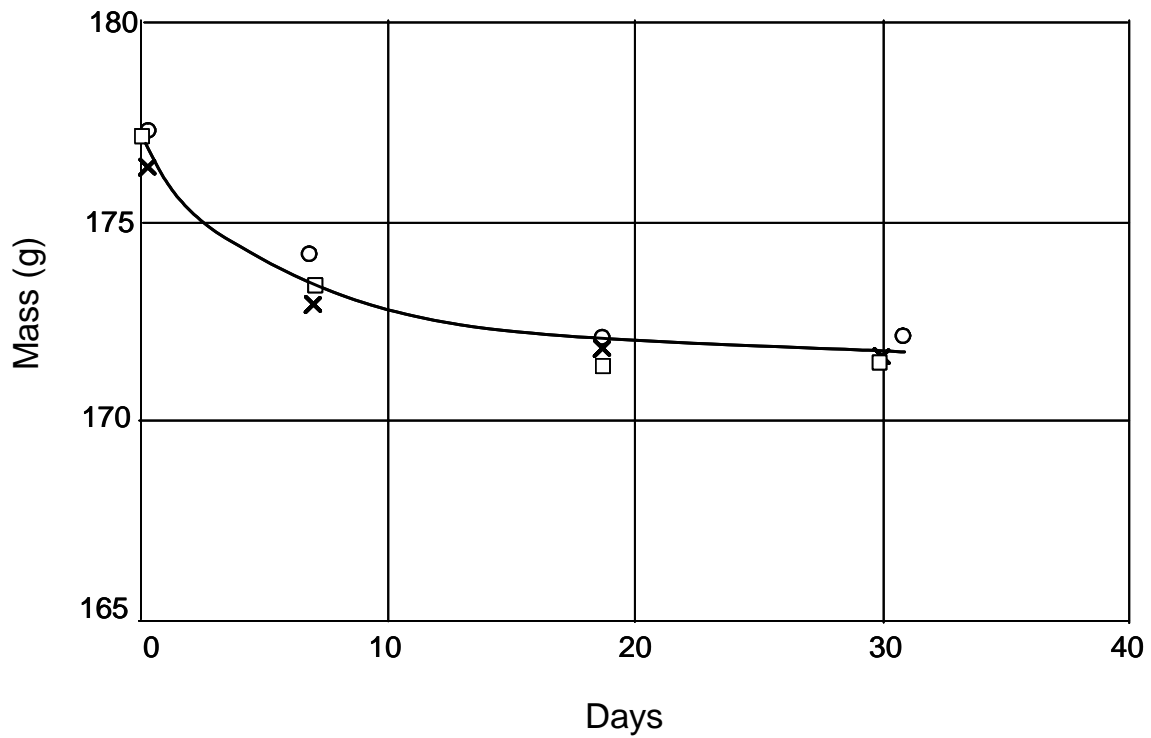

Figure 24 : Time to suction equilibration in triaxial samples submitted to a 800 kPa suction (Suraj De Silva 1987)

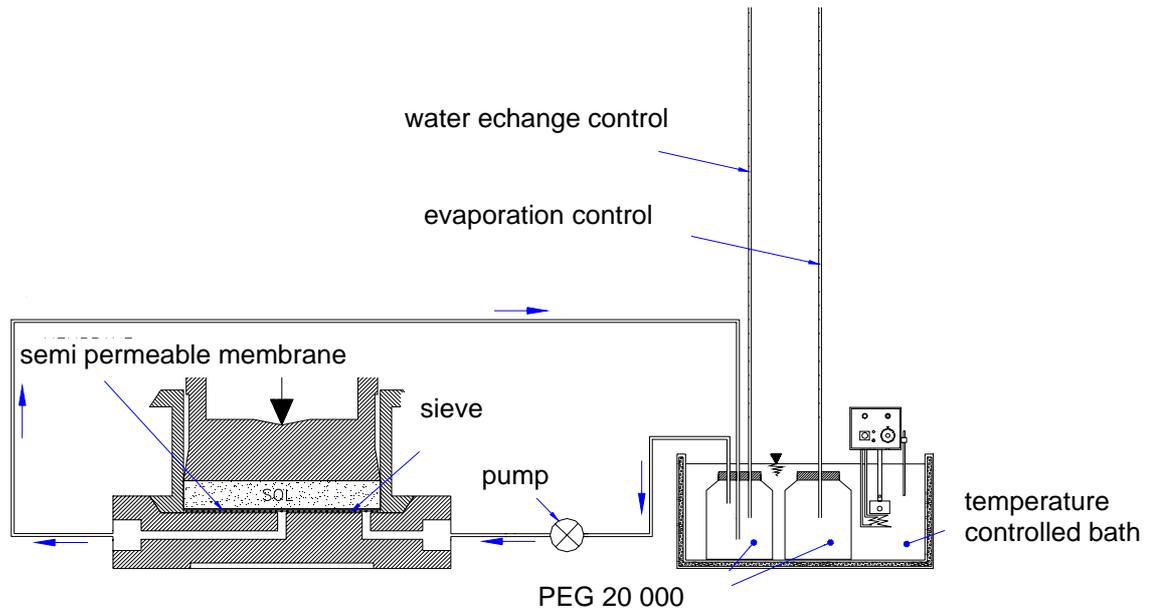

Figure 25 : Adaptation of a closed circuit to the osmotic oedometer (modified after Delage et al. 1992)

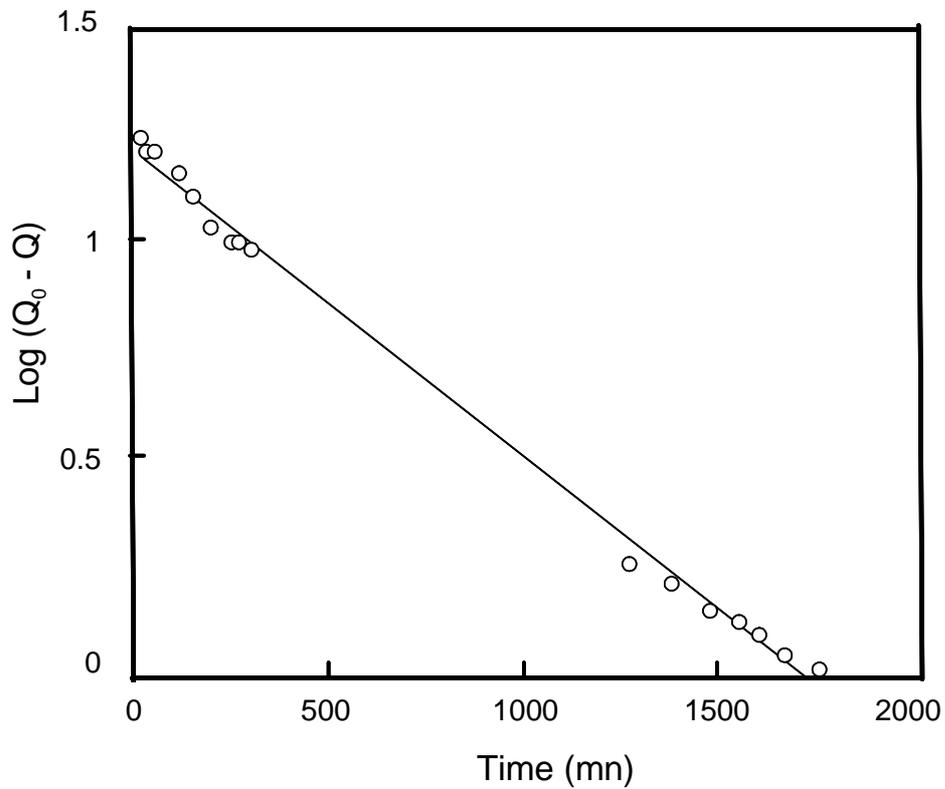

Figure 26: Adaptation of Gardner's method of measuring the permeability of unsaturated soils (Vicol 1990, Delage et al. 1992).

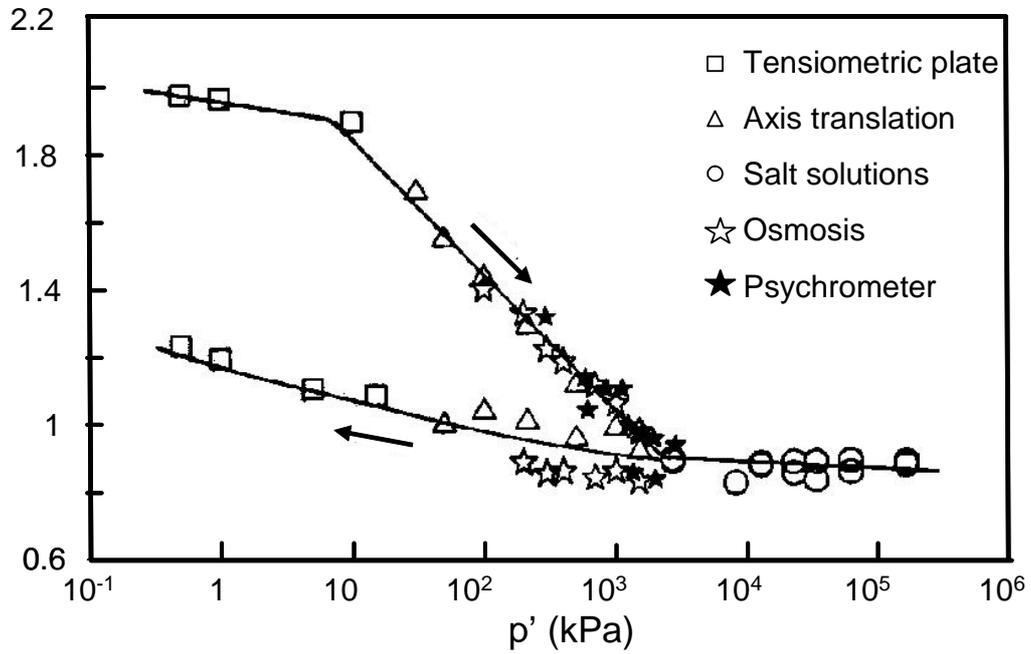

Figure 27 : Comparison between various suction controlled techniques (Fleureau et al. 1993)

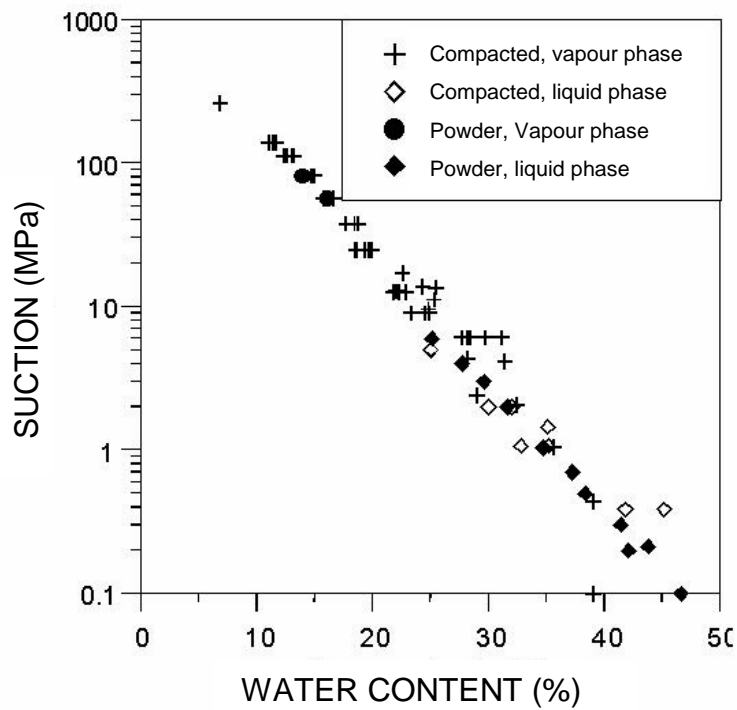

Figure 28 : Water retention properties of a engineered barrier FoCa7 clay (Yahia-Aissa et al. 2000)

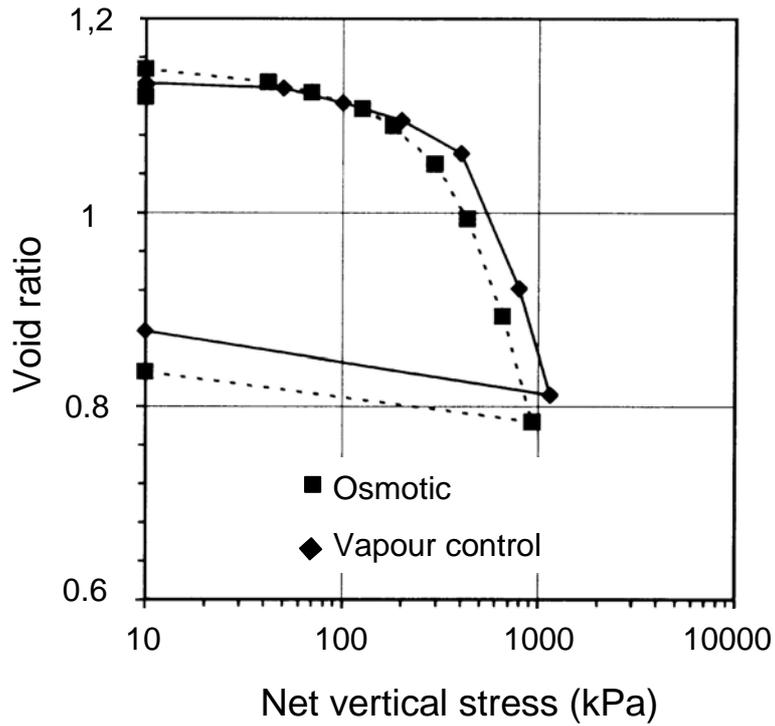

Figure 29 : High suction (8.5 MPa) oedometer compression tests (Cuisinier and Masrouri 2005)

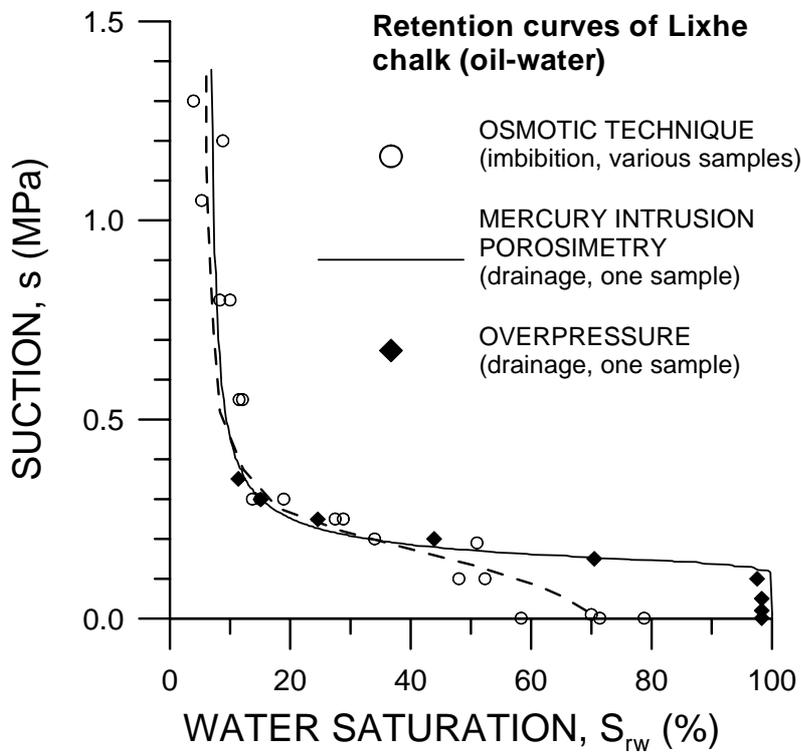

Figure 30 : Oil retention properties of a chalk (Priol et al. 2004)